# Helping Software Developers through Offline Repository Based API Searching in Data Mining Integrated Environment

Md. Ratul Uddin Ashraf[*], Anujoy Das[*], Ziaur Rahman[*], Ali Newaz Bahar[*], Husne Ara Rubaiyeat[γ]

[*]Department of Information and Communication Technology, [γ]Department of Computer Science

[*]Mawlana Bhashani Science and Technology University, Tangail-1902, [γ]National University, Gazipur-1704

Mail: ratulshourov01@gmail.com, anujoydas@gmail.com, zia@iut-dhaka.edu, bahar_mitdu@yahoo.com, rubaiyeat@yahoo.com

*Abstract*—Software development is getting changed so rapidly. It will be highly benefited if we can accelerate software development process by guiding developers. Appropriate guidelines and accurate recommendations to developers during development process can reduce software development expenses, as well as can save valuable times of developers. There are a number of approaches to speed up the software development process. It can be done through code assistance tools that help developers by recommending relevant items from searching particular repository of Application Programming Interface (API). Some approaches are based on online searching that have some drawbacks due to request and response latency as it has to deal with the extra-large files in a server. Developers generally uses previously completed resources as well as libraries or frameworks to generate relevant snippets which are supplied by the referral repository of APIs. Developers find it hard to choose the appropriate methods as there are thousands of methods in which some are not properly documented. In this paper we have proposed a concept and its respective framework to guide developers that suggests relevant API methods from an offline mined repository. From the investigation we made, we can say that our approach works much better than some of the existing approaches.

## I. INTRODUCTION

Software development processes for a system is basically done by writing a program based on specific design specification. Nowadays it can be done by using various techniques. To do this development process developers in the software industries generally creates new methods or modifies existing ones or reuses previously defined methods. One of the processes to develop software is the reuse of existing libraries and frameworks. Developers can be highly benefited if they get appropriate guidelines and accurate recommendations. It can decrease development cost and can save valuable time. Application Programming Interface (API) is a process to search relevant items from an existing repository. This process is widely used all over the world. Some approaches of these API search processes are based on online searching. Developers usually searches for API usage pattern using online repositories provided by different tools.

General code search engines often deals with the extra-large files. These extra-large files often contain a large number of API methods. Developers find it very difficult to choose the appropriate API methods to use as there are also duplicate methods. The API methods returned by the search engines are not mined so relevant data cannot be displayed. As the database of the search engines contains a large amount of data; it takes a long time to find appropriate methods. The online search processes for API methods are fully dependent on the server; as it is fully internet based it takes more time to generate the entire API methods thus making it a time consuming process. Normal API library or framework written in object oriented languages and it contains a huge number of methods and classes. If the API methods and classes do not have the meaningful name sometimes even experienced programmers often face a lot of problems. The API methods returned by the search engines are not often well documented. Developers find it challenging to use the most effective and appropriate methods.

To solve these problems we have proposed ORBAS (Offline Repository Based API Searching) thatis an offline repository that contains various API attributes. ORBAS mines API pattern usage methods from the innovative sequences of API in source code. The mined patterns contain less redundant patterns so that the developers cannot find the similar patterns repeatedly. The high coverage usage of API methods should be returned by the patterns to ensure a beneficial achievement to the developers.

In this paper we first plan to make a query in the repository for searching methods. Then the repository returns the associated source files of the matched inquiries. After that there is a code analyzer which extracts API methods from the source files. In the paper we have propose PMD [1] source code analyzer which extracts the API methods from the source code even if the source code is not compilable. The methods are then clustered into groups for mining. Then mining is implemented on the clusters to produce frequent closed sequence. We propose the BIDE [2] algorithm for mining frequently closed API usage patterns. We have again used a clustering technique in the frequent closed sequences. Thus two-step clustering techniques are implemented before and after the BIDE algorithm to produce more accurate usage patterns. Finally the closed API usage clusters are then produced and the related API usage methods are recommended to the developers. We have come to see that after following this approach ours are more efficient than some of the exiting approach to guide developers for getting API usage patterns.

## II. BACKGROUND AND RELATED WORKS

There exist a number of API usage pattern search theories or tools to find out useful API patterns from source code. Some of these are MAPO[3], MAC[4], ESDP[5], Hipikat[6], ParseWeb[7] etc. Often this search tools are based on online code search repositories such as Koders[8] and Google code search[9]. These code search engines generally evaluate code snippets containing keywords of API methods. Strathcoma[12] is a code snippet recommender, which finds out a set of code snippets by matching the structure of the codes. MAPO is a tool which mines API usage patterns and uses mined patterns for recommending code snippets. MAPO mines API usage pattern which have sequential information among the method calls. But the sequences patterns are not clustered in MAPO. ParseWeb accepts queries from the programmer and generates code samples based on the object type of the source and destination. UP-miner is also a mining tool for mining closed sequences returned from the API methods. It uses two-step clustering technique to find out frequently closed sequences. But it does not use a code analyzer to extract source files from the given result provided by search engines based on the programmer query.

Unlike many related works, we have developed ORBAS based on the mining algorithm to mine API methods and a two-step clustering technique to produce frequently closed sequences. As our process searches an offline repository we aim to make it a time consuming procedure.

## III. FRAMEWORK

To mine API usage patterns from an offline based repository we have developed a framework based on the mining and clustering of frequent closed sequences acquired from the source files which are resulted from the query performed in the repository. There are five major components in the framework: an offline repository, a code analyzer, sequence miner, sequence cluster and recommender. The implementation of ORBAS framework is given here. The total procedure works as illustrated in Fig. 1.

Step-1: In the ORBAS framework there is an offline repository in which programmer performs a query.

Step-2: A set of source code is returned from which the code analyzer extracts API methods.

Step-3: A two-step clustering technique is performed before and after mining API sequences.

Step-4: The best API usage cluster is produced and the required methods are recommended to the programmers.

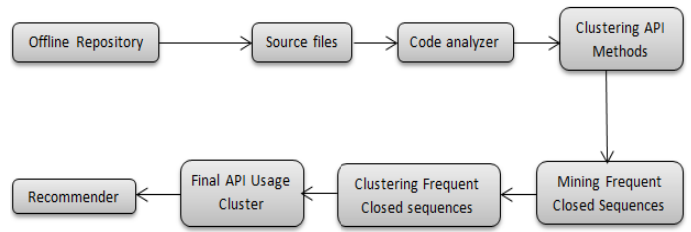

Fig. 1. An overview of ORBAS framework.

## IV. WORKING APPROACH

### A. Offline repository:

We have developed an offline repository in which different API methods are being inserted. The repository is organized by using a XML document to store API methods. The repository consists of sections such as search, update and delete. In the search section API methods can be searched by providing the method name. The update section makes necessary steps to add a new API method in the repository. It provides the insertion of methods. The delete section can remove API methods from the repository.

### B. Source files:

As our framework is based on offline repository so the source files are collected directly from the repository based on the query implemented in the repositories. The repository provides the API methods which matches the input query. If the input method matches with the API methods stored in the repository, the complete source files of the methods are returned by the query.

### C. Code analyzer:

Code analyzer helps us to extract the API methods from the source files. PMD [1] is a source code analyzer which does not need compilable source files for extracting API methods. PMD search outs common programming flaws like unused variables, empty catch blocks, unnecessary object creations etc. The code analyzer extracts a list of methods for each source files. This is done by collecting class name and full package name for each method. Based on the receiver object name we track the field declarations and local declarations to know the class name of the method calls. For the case of the classes which are exported in the import statement; we keep tracks of import statements to get the full package name of these classes.

### D. Clustering API methods:

The method sequences returned by the source code analyzer can be two types. One is the frequently used API patterns which have large threshold support and another is less frequently used patterns which has minimum threshold. Clustering technique increases the chance of less frequently used sequence patterns to mine out before mining the frequently closed sequence. *SeqSim* [10] is an n-gram based technique which computes the similarity of two different sequences.

The set G(X) is defined as the n-gram set for a sequence $S(s_1 s_2 \ldots s_n)$ as a collection of unigrams, bi-grams, … , n-gram of *S*:

(S) = {$s_1, s_2\ldots, s_n, s_1s_2, s_2s_3 \ldots, s_{n-1}s_n, \ldots, s_1s_2\ldots s_{n-1}, s_2s_3\ldots s_{n-1}s_n, s_1s_2\ldots s_{n-1}s_n$}

Two situations are considered to evaluate a similarity between two sequences:

i) Sequences are similar if they both equally have a large number of API methods between their representative n-gram sets.

ii) Sequences are similar if they both equally have common subsequences.

If two sequences are $s_1$ and $s_2$ then the similarities are:

$$\text{SeqSim}(s_1, s_2) = \frac{\sum_i \text{Weight}(g_{i\cap})}{\sum_i \text{Weight}(g_{iU})}$$

Where $g^i_\cap \in G_\cap = G(s_1) \cap G(s_2)$, $g^i_U \in G_U = G(s_1) \cup G(s_2)$, weight ($g^i_\cap$) is equal to the length of $g^i_\cap$.

As an example, given $s_1$ = abc and $s_2$ = cab, then we will have

$G_\cap$ = {a,b,c,ab}, $G_U$ = {a,b,c,ab,bc,ca,abc,cab} and Weight(ab) = 1, Weight(ab) = 2, SeqSim($s_1,s_2$) = 0.33.

Based on the result of SeqSim, we cluster the API sequences. The remoteness of two clusters is considered as the highest value of remoteness between a pair of items in these two clusters.

### E. Mining frequent closed sequence:

BIDE algorithm [11] is used to recognize the frequent API usages for mining frequent closed sequences. A sequence $s_1$ is a sub-sequence of $s_2$ and $s_2$ is a super-sequence of $s_1$ when $s_1$ is contained in $s_2$ and s is a frequent sequence. As there are no appropriate super sequences of s with the same support then s is a frequent sequence in a set of sequence if sup(s) ≥ min_sup. Sup(s) is called the support of sequence which is derived by the ratio of the sequences that are super sequences of s in the sequence set. Min_sup is the minimum threshold support [2]. To prepare the frequent closed sequences for the cluster, BIDE algorithm is applied to each cluster originated from the procedure described in the previous section. For example, let us assume there are three sequences is a cluster which are xy, xyz, xyt. If the minimum threshold support is 0.5, BIDE algorithm gives the result of frequent closed sequences xy, where other sequence mining algorithms will provide the result x, y and xy. If xy is frequent then it can be easily noted that x and y will also be frequent. Thus x and y does not need to be enlisted. BIDE algorithm provides the sequences in which there are subsequences of similar support.

### F. Clustering frequent closed sequences:

The API method sequences that are clustered using the proposed SeqSim technique in previous section; can be of two types. One is the sequences which are not similar to each other and another is the sequences which are similar. The sequences that are not similar are partitions into clusters. But the sequences that are similar can produce redundancy in the final API usage patterns. To solve this problem, another clustering is performed for the similar sequences. In this process the resulting clusters are considered as usage patterns. For example, we consider two frequent closed sequence xyz and zy which are mined from two distinct clusters. These two sequences are grouped together and shown as xyz after performing the clustering technique. The total number of superfluous sequences can also be removed using the technique. By applying the SeqSim technique; we can measure similarities between two frequent closed sequences.

### G. Recommender:

After performing the clustering techniques the frequent closed sequences are produced and the results are shown to the programmers. The programmers can bypass the process of checking code sequences one by one as the recommender returns the mined patterns as an index of sequences. The best API patterns are shown on the top of the index thus providing the programmers a suggestion to select which one is to be used.

## V. EVALUATION

### A. Environmental Setup:

The implementation of our work requires the usage of severe environmental setups. ORBAS, MAPO and Koders are used as search engine. To produce a graph based result we have implemented our result data in MATLAB. The system platforms in which we developed our framework contain core-i5 processor (2.30 GHz), RAM of 4GB and 3G network.

### B. Time Response Evaluation:

Our work in ORBAS is compared to the existing MAPO and Koders search engine for evaluating time response. For example we have consideredWebRequest.create as sequence 1, then we have considered another sequence HTTPWebRequest.getResponse as sequence 2 for both Table I and Table II and so on. By inputting these sequences we generate an output based on the time response of the search tools which is shown in Table I.

TABLE I. TIME RESPONSE EVALUATION

| Squ. No. | API Name | MAPO[3] | Koders[8] | ORBAS |
|---|---|---|---|---|
| 1 | WebRequest.create | 5 | 7 | 3 |
| 2 | HTTPWebRequest.getResponse | 7 | 9 | 2 |
| 3 | HTTPWebRequest.create | 6 | 7 | 4 |
| 4 | SQLConnection.open | 3 | 5 | 1 |
| 5 | SQLDatabase.read | 4 | 6 | 2 |
| 6 | SQLConnection.close | 2 | 5 | 1 |
| 7 | SQLCommand.parametersAdd | 4 | 8 | 2 |
| 8 | SQLDataAdapter.Bill | 5 | 7 | 1 |
| 9 | HTTPResponse.setContent | 6 | 5 | 3 |
| 10 | HTTPResponse.write | 7 | 5 | 3 |

Fig. 2 shows the graphical representation of the response time from the data of Table I. The graph shows result analysis of the time response evaluation of API sequences among ORBAS, MAPO and Koders. The graph is designed in MATLAB by inserting the values of Table I. From the graph it is clear that the time responses for each method are lesser in ORBAS than MAPO and Koders.

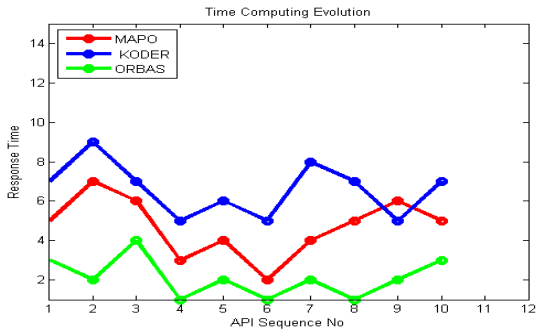

Fig. 2. Response time with different API sequences.

*C. Performance Evaluation:*

By inputting the sequences into different search engines we have evaluated the performance. Table II shows the performance evaluation based on the methods returned by each search tools. It is shown that ORBAS returns a few number of effective and useful methods compared to MAPO and Koders.

TABLE II. PERFORMANCE EVALUATION

| Squ. no. | API Name | MAPO [3] | Koders [8] | ORBAS |
|---|---|---|---|---|
| 1 | WebRequest.create | 71 | 110 | 8 |
| 2 | HTTPWebRequest.GetResponse | 58 | 95 | 7 |
| 3 | HTTPWebRequest.create | 22 | 55 | 4 |
| 4 | SQLConnection.open | 94 | 105 | 12 |
| 5 | SQLDataReader.Read | 67 | 82 | 10 |
| 6 | SQLConnection.close | 44 | 50 | 2 |
| 7 | SQLCommad.parametersAdd | 56 | 70 | 8 |
| 8 | SQLDataAdapter.Bill | 19 | 25 | 3 |
| 9 | HTTPResponse.setcontent | 8 | 14 | 2 |
| 10 | HTTPResponse.write | 5 | 9 | 1 |

Fig. 3 shows the graphical representation of the number of methods returned by per API sequence. The graph shows the result analysis of the number of methods returned for ORBAS, MAPO and Koders. The graph is designed in MATLAB by inserting the data of Table II. It is clear from the graph that ORBAS returns a few but useful API methods than MAPO and Koders.

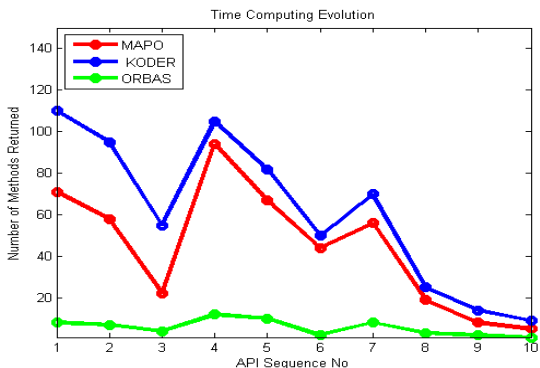

Fig. 3. Methods return from different API sequences.

## VI. CONCLUSION AND FUTURE WORK

With a view to helping developers to understand API usages and client code more nicely we have developed a framework called ORBAS which is a supporting tool for mining API usages from anoffline repository. Our experimental results show that ORBAS generates useful API pattern than the existing tools. In future we will try to focus on automatic code generation based on the returned API patterns. Then these automatic generated codes can easily be inserted into developer's code. Our framework ORBAS may require some more data mining techniques. We further need to investigate to confirm whether the selected technique is the best or not. For the purpose of mining API usage patterns we plan to apply other clustering techniques such as k-means clustering. We also have a plan to apply other partial order miner in the mining stage. We plan to include other features such as class structure for clustering. We plan to find out the weights and thresholds value. We will concern other approach and technique and compare with our current process. In our working procedure we do not see the quality of the mined pattern directly. Usually most libraries do not provide usage pattern, there islacking in the field of the usage pattern. We must concern to experiment the quality of the usage pattern. Our framework approach must be applicable to many other object oriented languages. We will try to adapt our supporting tool in other large-scale databases. In our approach we have extracted API usages only for the API client code. In the future we will try to extract to the client code and the implementation code both.